\documentclass[12pt,twoside]{article}
\usepackage{fleqn,espcrc1}
\usepackage{graphicx,epsfig}

\newcommand{\AmS}{{\protect\the\textfont2
  A\kern-.1667em\lower.5ex\hbox{M}\kern-.125emS}}

\newcommand{\gd}{\delta}

\begin{document}
\renewcommand{\textfraction}{0.0001}
\renewcommand{\bottomfraction}{0.999}
\renewcommand{\topfraction}{0.9995}

\title{Formation of Few-Body Clusters in Nuclear Matter}

\author{M. Beyer, FB Physik, Universit\"at Rostock, 18051 Rostock,
  Germany}

\maketitle

To describe the formation of clusters -- in the general case a
nonequilibrium process -- in an interacting many-body system
constitutes a new challenge for few-body methods.  This may happen
when the residual interaction between the quasi-particles leads to
correlations. An example for such a system is nuclear matter. In the
laboratory finite pieces of nuclear matter can be produced in heavy
ion collisions. In astrophysics nuclear matter occurs e.g. during the
supernova collapse and the formation of a neutron star.

A microscopic approach to treat the formation of clusters uses a
generalized quantum Boltzmann equation. This coupled equation has been
numerically solved for nucleon $f_N$, deuteron $f_d$, triton $f_t$,
and helium-3 $f_h$ distributions utilizing the
Boltzmann-Uehling-Uhlenbeck (BUU) approach~\cite{dan91}.  The coupling
between the different species is through the collision integrals
${\cal K}[f_N,f_d,\dots]$. For the deuteron loss ${\cal K}^{\rm
  out}_d(P,t)$, e.g.,  it is
\begin{eqnarray}
{\cal K}^{\rm out}_d(P,t)&=&
\int d^3k\int d^3k_1d^3k_2d^3k_3\;
|\langle k_1k_2k_3|U_0|kP\rangle|^2_{dN\rightarrow pnN}\nonumber\\&&
\qquad\times
\bar f_N(k_1,t)\bar f_N(k_2,t)\bar f_N(k_3,t)f_N(k,t)
+\dots\label{eqn:react2}
\end{eqnarray}
where $\bar f_N=(1-f_N)$. The ellipsis denote further possible
contributions, e.g. $dd\rightleftharpoons tp$, $dd\rightleftharpoons
hp$ or processes like $\gamma d\rightleftharpoons np$, etc.  The
quantity $U_0$ is the $Nd\rightarrow NNN$ break-up transition operator
that in general depends on the medium.  This dependence is neglected,
if {\em experimental cross sections} are used to replace $U_0$ in
Eq.~(\ref{eqn:react2}) -- a standard technique and in many cases very
successful. To calculate $U_0$ including the {\em self energy shift}
and the proper {\em Pauli blocking} and study the influence of the
medium on different observablesa generalized Alt-Grassberger-Sandhas
(AGS) equation~\cite{AGS} has been derived
earlier~\cite{bey96,bey97,beyFB,schadow,kuhrts,alpha}. The effective
few-body problem in matter arises in the Green function
approach~\cite{fet71} along with a cluster mean-field
expansion~\cite{RMSS} or the a Dyson equation approach~\cite{duk98}.

\begin{figure}[tb]
\begin{minipage}{0.45\textwidth}
\epsfig{figure=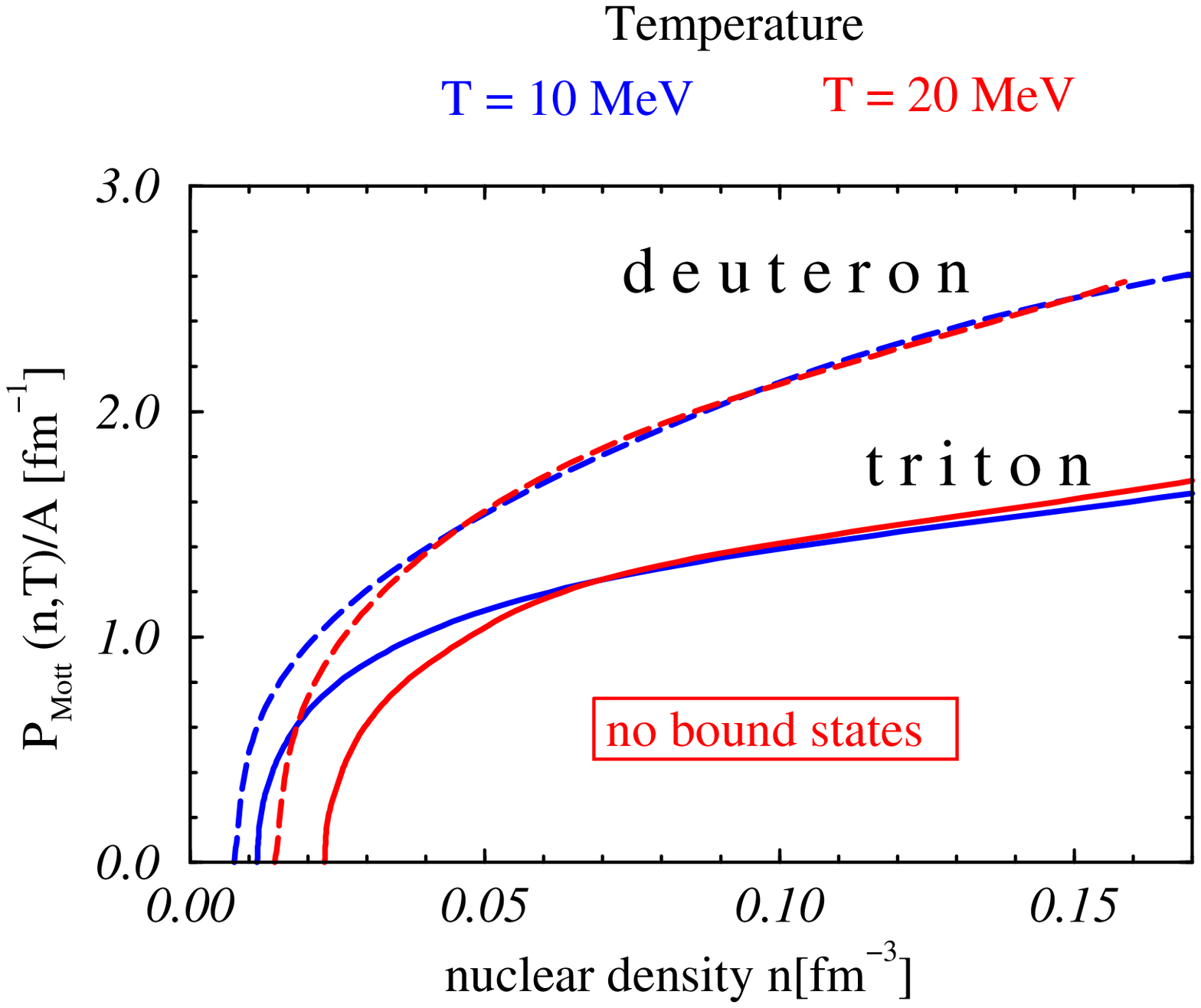,width=\textwidth}
\vspace*{-1.5cm}
\caption{\label{fig:mott} 
  Mott momenta for triton and deuteron.}
\end{minipage}
\hfill
\begin{minipage}{0.45\textwidth}
\epsfig{figure=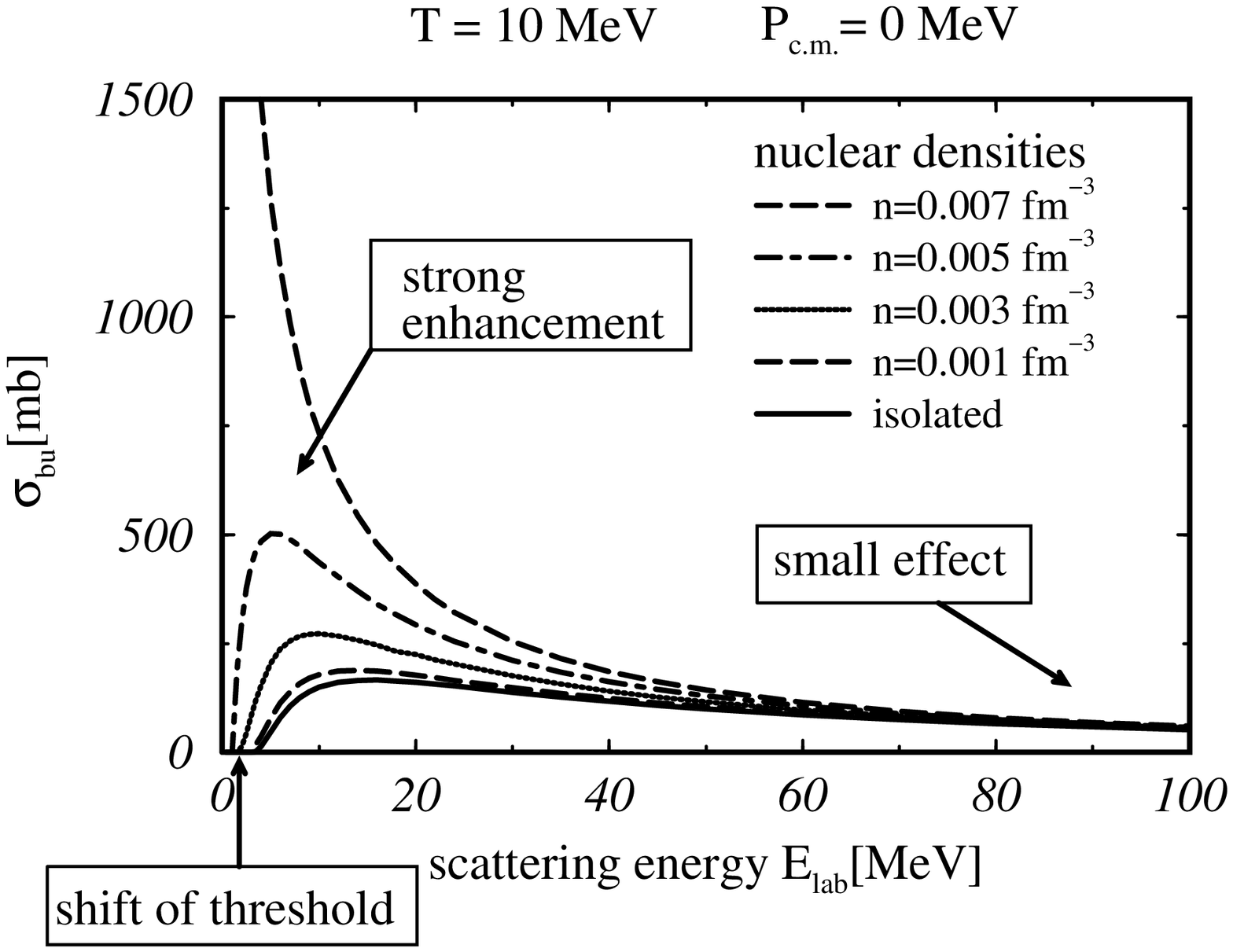,width=\textwidth}
\vspace*{-0.7cm}
\caption{\label{fig:breakup} 
$Nd\rightarrow NNN$ break-up cross section for different
  densities.}
\end{minipage}
\end{figure}
\begin{figure}[b]
\begin{minipage}{0.45\textwidth}
\epsfig{figure=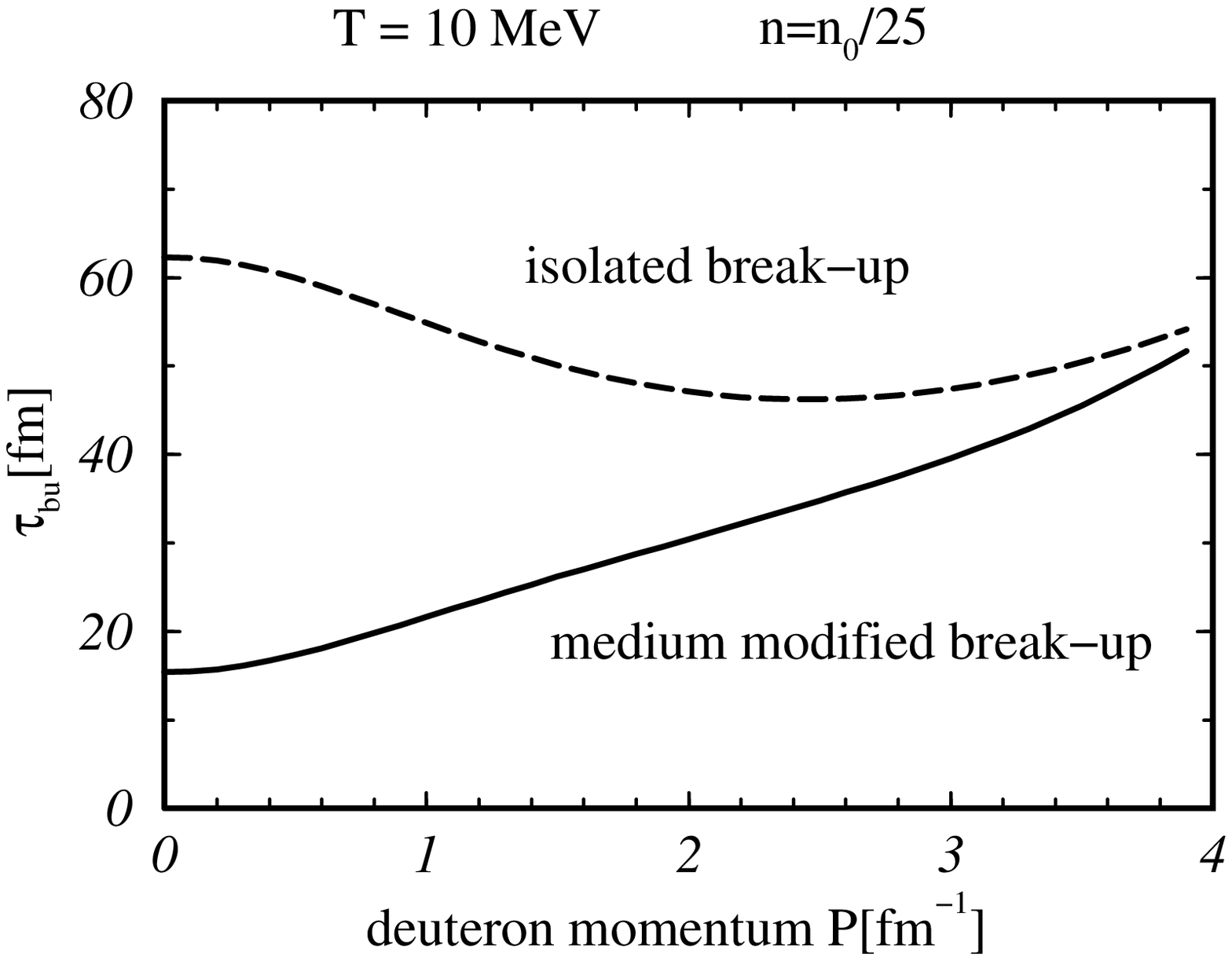,width=\textwidth}
\vspace*{-1.4cm}
\caption{\label{fig:life} Deuteron break-up time as a function of the
  deuteron momentum.}
\end{minipage}
\hfill
\begin{minipage}{0.45\textwidth}
\epsfig{figure=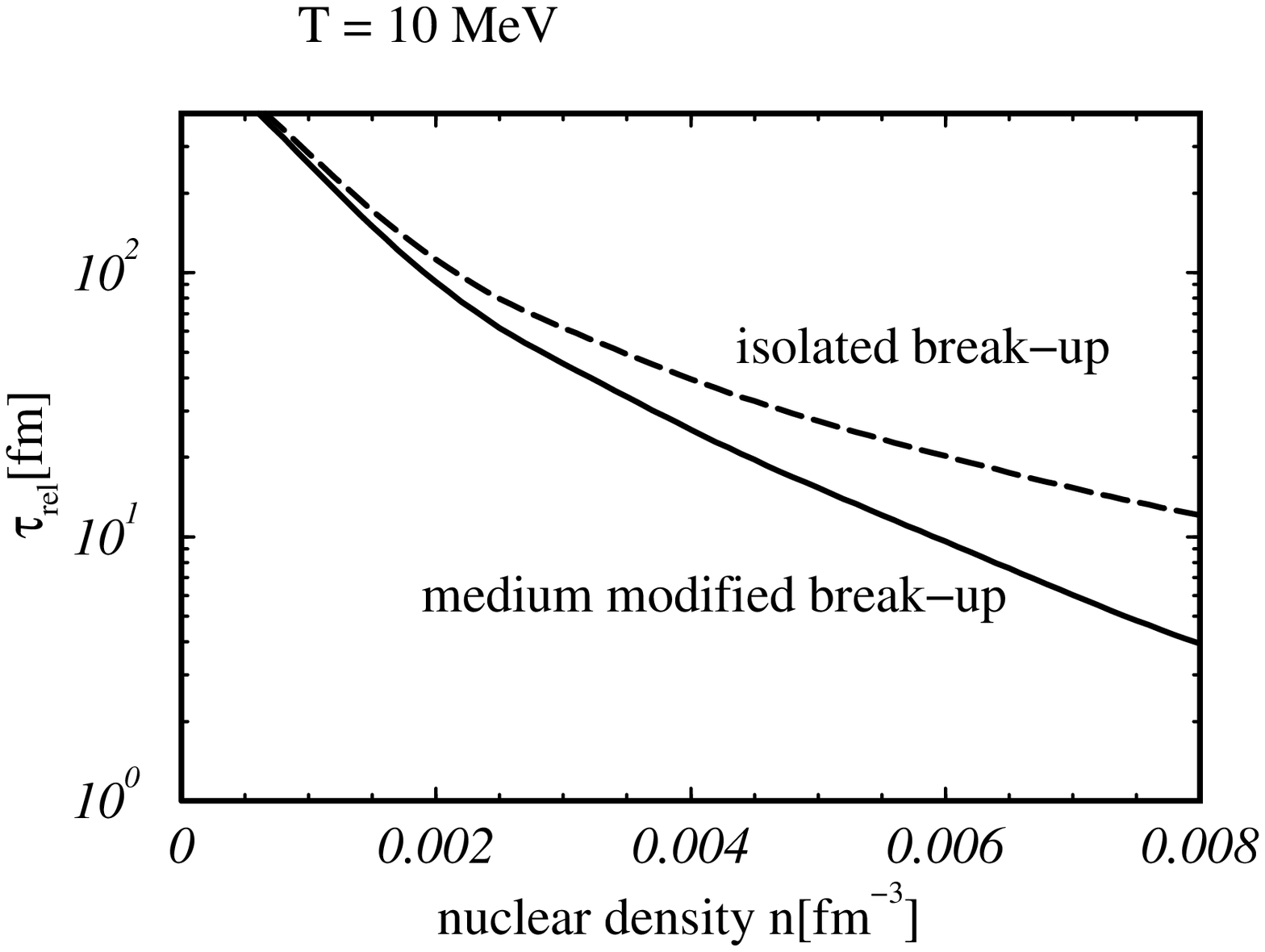,width=\textwidth}
\vspace*{-1cm}
\caption{\label{fig:chem} Chemical relaxation time as a
  function of nuclear density.}
\end{minipage}
\end{figure}

Using the effective equations derived
elsewhere~\cite{bey96,bey97,beyFB,schadow,kuhrts,alpha} we may study
two important effects of the medium on the effective few-body-systems
embedded in nuclear matter: (1) The change of binding energy, i.e. the
self energy shift, (2) the changes in the reactions rates. Both
effects are important and have consequences for the simulation of
heavy ion collisions.

The change of binding energy eventually leads to the Mott effect
(where $E_{\rm bound} \rightarrow 0^-$). Not as dramatic as in
Coulombic systems -- here the Mott effect leads to the transition from
isolating to conducting phase -- it, however, influences the number of
clusters and the energy spectrum produced in a heavy ion collision.
The Mott density depends on the momentum of the cluster as for higher
momenta blocking of the constituents of the cluster is less effective.
The Mott momenta for deuteron and triton are shown in
Fig.~\ref{fig:mott} for two different temperatures. Tritons are more
stable than deuterons and at lower densities both clusters a more stable
for higher temperatures.

For a typical temperature of the heavy ion collision (in the final
stage) we have calculated the {in-medium} cross
section. The result is shown in Fig.~\ref{fig:breakup}. The threshold
shift is because of the change of the deuteron's binding energy.  A
strong enhancement of the maximum of the cross section appears,
however less change at higher energies.

As a consequence the reaction time scales become much faster, when
{\em in-medium} rates are used in the calculation instead of {\em
  isolated} ones.  Fig.~\ref{fig:life} shown the deuteron break-up
time ${\tau_{\rm bu}(P,n_N)}$ evaluated in linear response, where the
life time of deuteron fluctuations $\gd f_d(P,t)$  depends on
the deuteron momentum $P$ and the nuclear density~\cite{bey97}.
Similar, a chemical relaxation time ${\tau_{\rm rel}(n_N)}$ can be
defined, which results from a linearization of the respective rate
equations~\cite{kuhrts}.  The relaxation times are shown in
Fig.~\ref{fig:chem}.

\begin{figure}[tb]
\begin{minipage}{0.45\textwidth}
\epsfig{figure=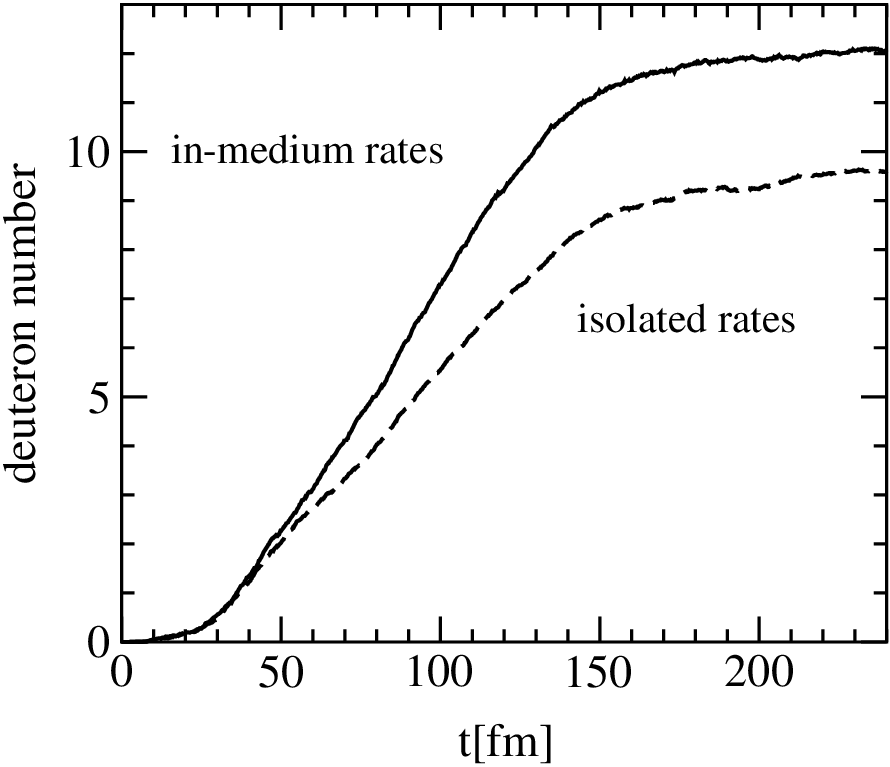,width=\textwidth}
\vspace*{-1.2cm}
\caption{\label{fig:PLT} 
  Total integrated number of deuterons as a function elapsed time.}
\end{minipage}
\hfill
\begin{minipage}{0.45\textwidth}
\epsfig{figure=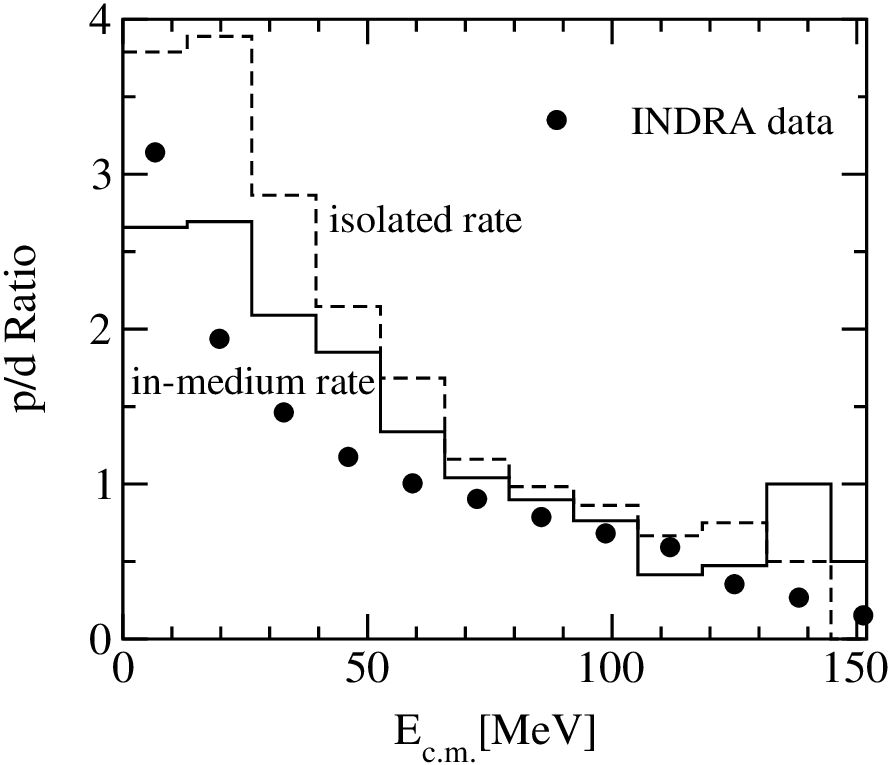,width=\textwidth}
\vspace*{-1.2cm}
\caption{\label{fig:pd2} 
  Ratio of proton to deuteron numbers as a function of c.m. energy.}
\end{minipage}
\end{figure}

Finally considering a specific heavy ion collision it is possible to
calculate the total number of deuterons coming out of a central
collision of $^{129}$Xe on $^{119}$Sn at 50 MeV/A~\cite{INDRA}.
Fig.~\ref{fig:PLT} shows the integrated number of deuterons taking
into account gain and loss terms induced by the reactions in the
medium. The net effect is a significant enhancement of the number of
deuterons.  Fig.~\ref{fig:pd2} shows the influence of using {\em
  in-medium} rates on the spectrum of the proton to deuteron ratio
along wuth experimental data.

From the analysis is becomes clear that the {\em medium modified
  elementary} cross section and the proper {\em self energy
  correction} (binding energy shift, Mott effect) of the clusters
should be included in the simulation of heavy ion collisions at that
energies.

{\em Acknowledgment:} It a pleasure for me to thank P.  Danielewicz,
C. Kuhrts, G. R\"opke, W.  Schadow, and P. Schuck for substantial
contributions. This work and presentation has been supported  by the
Deutsche Forschungsgemeinschaft.

\end{document}